\documentclass[12pt,reqno]{amsart}
\usepackage[foot]{amsaddr}
\usepackage{graphicx,amsmath,amssymb,mathtools,amsthm}
\usepackage{color}
\usepackage{natbib}
\usepackage{float}
\usepackage{bbm}
\usepackage{longtable}
\usepackage{caption}
\usepackage{subcaption}
\usepackage{enumitem}
\usepackage{appendix}

\usepackage{xr}
\usepackage{xr-hyper}
\usepackage{hyperref}
\hypersetup{colorlinks=true,
citecolor = blue,
filecolor=cyan}
\usepackage{cleveref}

\usepackage[left=2cm,right=2cm, top=2cm, bottom=2cm]{geometry}
\long\def\comment#1{}

\newtheorem{theorem}{Theorem}

\newtheorem{lemma}{Lemma}

\theoremstyle{definition}

\newtheorem{remark}{Remark}

\title{A maximal inequality for local empirical processes under weak dependence}

\author[Luis Alvarez]{Luis A. F. Alvarez}
\address{Alvarez: Post-doctoral researcher, Fundação Getúlio Vargas.}
\author[Cristine Pinto]{Cristine C. X. Pinto}
\address{Pinto: Full professor, Insper.}

\date{June 2023}
\begin{document}

	\maketitle
	\begin{abstract}
		We introduce a maximal inequality for a local empirical process under strongly mixing data. Local empirical processes are defined as the (local) averages $\frac{1}{nh}\sum_{i=1}^n \mathbf{1}\{x - h \leq X_i \leq x+h\}f(Z_i)$, where $f$ belongs to a class of functions, $x \in \mathbb{R}$ and $h > 0$ is a bandwidth. Our nonasymptotic bounds control estimation error uniformly over the function class, evaluation point $x$ and bandwidth $h$. They are also general enough to accomodate function classes whose complexity increases with $n$. As an application, we apply our bounds to function classes that exhibit polynomial decay in their uniform covering numbers. When specialized to the problem of kernel density estimation, our bounds reveal that, under weak dependence with exponential decay, these estimators achieve the same (up to a logarithmic factor) sharp uniform-in-bandwidth rates derived in the iid setting by \cite{Einmahl2005}.
		\end{abstract}
	
		\section{Introduction}
		
		In this paper, we introduce a maximal inequality for a local empirical process under exponential decay of the sample $\alpha$-mixing coefficients. We provide a nonasymptotic bound on an Orlicz norm of the uniform estimation error of the local-average process $\frac{1}{nh}\sum_{i=1}^n \mathbf{1}\{x - h \leq X_i \leq x+h\}f(Z_i) $, where uniformity holds simultaneously over the evaluation point $x \in \mathbb{R}$, bandwidth $ h \in [a_n, b_n)$, and function $f  \in \mathcal{F}$, with $a_n \leq b_n$ being positive constants and $\mathcal{F}$ a function class. The nonasymptotic nature of our results allows for function classes whose complexity, as measured by their uniform covering numbers, increases with $n$. The latter is especially useful for applications in high-dimensional statistics (see \cite{Belloni2017} for an example). We also discuss how to extend our results to multidimensional $x$ and subgeometric decay of mixing coefficients.
		
		We then apply our results to functional classes $\mathcal{F}$ that exhibit polynomial decay in their uniform covering numbers. This class is particulary interesting, as polynomial decay in uniform entropy is ensured by a finite VC dimension \citep[Theorem 2.6.7]{vanderVaart1996}. When specialized to the problem of kernel density estimation, our results show that, under exponential decay in the mixing coefficients, kernel estimators achieve the same, up to a logarithmic factor, sharp uniform-in-bandwidth rates obtained by \cite{Einmahl2005} in the iid setting.

		To the best of our knowledge, the closest paper to ours is \cite{Escanciano2020}, who provides uniform-in-bandwidth asymptotic rates for local empirical processes under stationarity, a $\beta$-mixing condition and polynomial decay in bracketing entropy. There are some important differences between his approach and ours, though. First, \citeauthor{Escanciano2020}'s analysis is asymptotic, thus not applicable to function classes of increasing complexity. Second, while the author uses bracketing entropy to control complexity, our analysis relies on uniform covering numbers. Finally, his main result (Theorem 2.1) leads to a rate of $a_n^{-1/2}$ over the function classes encompassed by his setting. When compared to our application in \Cref{sec_app}, we see that our resulting rates are logarithmic in $a_n$, thus offering a great improvement over his results in kernel-type problems.
			
			The remainder of this paper is organized as follows. Section \ref{sec_ineq} introduces our main result. Section \ref{sec_app} applies it to function classes with uniform entropy displaying polynomial decay. Section \ref{sec_conc} concludes.
			
	\section{A maximal inequality for local empirical processes}
	\label{sec_ineq}
	We present our main result in the theorem below. In what follows, we define, for some $p \geq 0$, $\psi_p(x) \coloneqq \exp(-x^p) - 1$; and, for a random variable $Z$, the Orlicz norm with respect to $\psi_p$ is given by:
	
	$$\lVert Z \rVert_{\psi_p} \coloneqq \inf \left\{C > 0: \mathbb{E}\left[\psi_p\left(\frac{|Z|}{C}\right)\right]<\infty\right\} \, .$$
	
	Moreover, for a sequence $\{Y_i\}_{i=1}^n$ of random variables defined on a common probability space $(\Omega, \Sigma, \mathbb{P})$, we define the $\alpha$-mixing coefficients as, for each $k \in \mathbb{N}$.
	
	$$\alpha(k) \coloneqq \sup_{t \in \mathbb{N}} \sup_{A \in \sigma(Y_1,\ldots, Y_t), B \in \sigma( Y_{t+k},Y_{t+k+1},\ldots)} |\mathbb{P}[A\cap B] - \mathbb{P}[A]\mathbb{P}[B]| \, .$$
		\begin{theorem}
			\label{theorem_main}
		Let $\{(X_i, Z_i)\}_{i \in \mathbb{N}}$ be a sequence of random variables defined on a common probability space $(\Omega, \Sigma, \mathbb{P})$, with mixing coefficients satisfying, for some $c > 0$, $\alpha(i) \leq \exp(-2 c i)$ $\forall i \in \mathbb{N}$. Suppose that each $X_i$ is real valued, with common distribution fucntion $G$ that admits a bounded Lebesgue density $g$. Suppose that each $Z_i$ takes values on a measurable space $(\mathcal{Z}, \Sigma)$.  Let $\mathcal{F}$ be a pointwise measurable class of real-valued functions with domain $\mathcal{Z}$. Define:
		
		$$S_n(x,f; h) \coloneqq \frac{1}{\sqrt{nh}}\sum_{i=1}^n \mathbf{1}\{ x - h \leq X_i \leq x+h\} f(Z_i) \, , \quad u \in [0,1],  f \in \mathcal{F}, h \in [a_n,b_n) \, ,$$
		where $0<a_n\leq b_n $ are positive constants, with $ \frac{1}{\lVert g\rVert_\infty n } \leq b_n < \frac{1}{\lVert g\rVert_\infty }$.
		
		Suppose that the class of functions $\mathcal{F}$ admits a measurable and  bounded envelope $\kappa$. Assume that, for some $q \in (1,\infty]$, $\omega(\epsilon) \coloneqq \sup_{P \in \mathcal{P}(\mathcal{Z})}\mathcal{N}(2\epsilon \lVert \kappa \rVert_\infty,\mathcal{F}, \lVert \cdot \rVert_{q,P})$, is finite for every $\epsilon > 0$, where $\lVert f \rVert_{P,q} = (P|f|^q)^{1/q}$ and the supremum is taken over all probability measures on $(\mathcal{Z}, \Sigma)$. Then, for every $n \geq 2$:
		\begin{equation}
			\label{eq_bound_thm}
			\left\lVert \sup_{x \in \mathbb{R}, h \in [a_n, b_n), f \in \mathcal{F} }|S_n(x,f;h) - \mathbb{E}[S_n(x,f;h)]| \right\rVert_{\psi_1}\leq C\frac{\lVert \kappa \rVert_\infty}{\sqrt{na_n}} + \frac{D \lVert \kappa \rVert_\infty}{\sqrt{na_n}} \sum_{l=\lceil -\log_2(\lVert g\rVert_\infty b_n)\rceil}^{\lceil\log_2(n) \rceil}  \inf_{\delta \geq 0} \psi_l(\delta) \, ,
		\end{equation}
		where $C>0$ is an absolute consant; the constant $D>0$ depends solely on $c$; and:
	\begin{equation}
		\begin{aligned}
				\psi_l(\delta) = \left(\log(n)^2[(l+1)\lor \log(\omega(\delta))] + \sqrt{\frac{ln}{2^l} \lor 1}\sqrt{[(l+1)\lor \log(\omega(\delta))]}\right) + \\
			n\delta \left[\left(n^{-1}\log(n)^2(l+1) + \sqrt{\frac{l}{2^l n} \lor \frac{1}{n^2}}\sqrt{(l+1)}\right)^\frac{{q-1}}{q} +\frac{1}{2^{\frac{(q-1)}{q}(l+1)}}\right] \, .
		\end{aligned}
	\end{equation}
	\end{theorem}
	
	The proof of Theorem \ref{theorem_main} is deferred to \Cref{app_proof_main}. It relies on a chaining argument due to \citeauthor{Rio2017} (\citeyear[Proposition 7.2]{Rio2017}), which we couple with the Bernstein inequality of \cite{Merlevede2009} to help achieve control of the error of the localizing function $\mathbf{1}\{x \leq X \leq x + h \}$ uniformly over the bandwidth. Note that our bound is nonasymptotic, with the constant $D$ depending solely on the decay parameter $c$. Our bound also depends on minimising the functions $\psi_l(\delta)$. This minimisation involves finding a ``small'' $\delta$ that is able to ensure appropriate control of the uniform entropy $\omega(\delta)$. In the next section, we show that, for classes with polynomial decay in their uniform entropy, a useul upper bound to this infimum can be computed.
		
		\begin{remark}[Extension to nid $X$]
			It is possible to extend our results to the setting where the distribution of $X_i$ varies with $i$. In this case, one has to adapt Lemma \ref{lemma_maximal} in Appendix \ref{app_proof_main} to allow for a different localization parameter $K_i$ for every $i$. Since this brings clutter to the proofs and does not reveal any new insights, for ease of exposition, we focus on the case where the $X_i$ are identically distributed. Observe that, in the statement of \Cref{theorem_main}, we do not require the $Z_i$ to be identically distributed.
		\end{remark}
		\begin{remark}[Extension to multidimensional $X$]
			It is possible to extend our results to the setting where $X$ is multidimensional by relying on Lemma 7.2 of \cite{Rio2017}. This result decomposes the difference between a $d$-dimensional cdf at two points as a sum of $d$ differences of the $d$ marginal cdfs at different evaluation points. Using this result, we may extend Lemma \ref{lemma_maximal} in Appendix \ref{app_proof_main} to the multivariate setting, and use it to prove a multivariate version of Theorem \ref{theorem_main}.
		\end{remark}
	\begin{remark}[Extension to unbounded envelopes and subgeometric decay]
		\label{rmk_extend_un}
		The statement of \Cref{theorem_main} requires the function class to display bounded envelope; and mixing coefficients to exhibit exponential (geometric) decay. These assumptions can be relaxed by replacing the Bernstein inequality of \cite{Merlevede2009} with a result due to \cite{Merlevede2010}. In this case, a tradeoff between the tail behaviour of the envelope function and the decay of mixing coefficients emerges. Consequently, we achieve control over the uniform estimation error under a different Orlicz norm. 
	\end{remark}

	\section{Application: function classes with polynomial decay}
	\label{sec_app}
	
	In this section, we apply our results to classes of functions that exhibit polynomial decay in their covering numbers. Specifically, in the setting of \Cref{theorem_main}, we take $q=2$ and consider a sequence of bounded function classes $\{\mathcal{F}_n\}_{n \in \mathbb{N}}$ with corresponding envelope $ \kappa_n$, whose uniform entropy numbers $\omega_n$ satisfy:
	
	$$\omega_n(\delta) \leq	 C_n \left(\frac{1}{\delta}\right)^{v_n} \,, \quad \forall \delta \in (0,1],$$
	for positive constants $C_n> 0$,$v_n > 0$, $n \in \mathbb{N}$. We also assume the envelopes to be uniformly bounded, with $\sup_{n \in \mathbb{N}} \lVert \kappa_n\rVert_\infty  \leq \phi < \infty$. Suppose that $na_n \to \infty$. In this case, we may find a constant $G>0$ such that, for every $n$ above a certain threshold, $ \lceil - \log_2(\lVert g \rVert_\infty b_n )\rceil \leq l \leq \lceil \log_2(n)\rceil$; and  $\delta < \exp\left(\frac{1}{v_n}(\log(C_n) - (l+1))\right)$:
	\begin{equation}
		\label{eq_upper}
		\begin{aligned}
			\psi_l(\delta) \leq G  \log(n)^2 (\log(C_n) - v_n \log(\delta)) + G \sqrt{\frac{l n}{2^l}} \sqrt{(\log(C_n) - v_n \log(\delta))} + G n \delta \frac{\log(n)^{3/2}}{2^{(1/2)(l+1)}}\,,
		\end{aligned}
	\end{equation}

	Consider the choice:
	
	$$\delta_l = \frac{C_n^{1/v_n}}{2^{((1/v_n))(l+1)}\sqrt{n} (\lVert g \rVert b_n)^{1/v_n} \log(n)^{3/2}}\, .$$
	
	Assuming that:
	
	$$\frac{\log(n)^3 \log \log n}{\sqrt{Ta_t}} \to 0 \, .$$

	In this case, we obtain the rate:

	\begin{equation}
		\begin{aligned}
					\left\lVert \sup_{x \in \mathbb{R}, h \in [a_n, b_n), f \in \mathcal{F} }|S_n(x,f;h) - \mathbb{E}[S_n(x,f;h)]| \right\rVert_{\psi_1}= \\ O\left(\sqrt{\frac{b_n}{a_n}}(C_n^{1/v_n}\lor \sqrt{-\log(\lVert g \rVert b_n)v_n \log(n)}\lor -\log(\lVert g \rVert b_n)) \right) \, .
		\end{aligned}
	\end{equation}
	
	Our rate explicitly accomodates for function classes of increasing complexity. If the sequence of functions has finite (albeit possibly increasing) VC dimension, then Theorem 2.6.7 of \cite{vanderVaart1996} shows that the $v_n$ and $C_n$ may be taken such that $C_n^{1/v_n}$ is bounded. In this case, the complexity of the function class only affects the rate through the $\sqrt{v_n}$ term, where $v_n$ scales linearly with the VC dimension.
	
	Finally, we consider the problem of kernel density estimation. In this case, the function class is the same for every $n \in \mathbb{N}$. If the bandwidths are taken such that $a_n \asymp b_n$, then our rate simplifies to:
	
	$$\sqrt{-\log(a_n)}\sqrt{\log(n) \lor  -\log(a_n)} \, .$$
	
	In contrast, Remark 2 in \cite{Einmahl2005} shows that, in the iid setting, one can achieve the rate:
	
	$$\sqrt{\log \log (n) \lor - \log(a_n)} \, ,$$
	which coincides with our rate except for logarithmic factors. Note that, if we consider, as it is usually done in practice, polynomial bandwidths, i.e $a_n = C n^{-\alpha}$ for $\alpha > 0$, then our rate simplifies to $\log(n)$, whereas \citeauthor{Einmahl2005}'s collapses to $\sqrt{\log(n)}$.

	\section{Concluding remarks}
	\label{sec_conc}
	
	In this paper, we introduced a maximal inequality for the uniform estimation error of a local empirical process under strongly mixing data, where uniformity holds simultaneously over the function class, bandwidth and evaluation point. Our nonasymptotic bounds accomodate function classes with increasing complexity, which is a useful feature for ``high-dimensional'' statistical analyses. As an application, we computed our bounds to function classes that exhibit polynomial decay in their uniform entropy. When specialized to the kernel density estimation problem, these results show that our bound leads to the same optimal rates derived by \cite{Einmahl2005} in the iid setting.
	
	More generally, we view our results as a first step in the development of rigorous uniform inference tools in local estimation problems under weak dependence and data-driven bandwidths. Specifically, one may combine our results with couplings in the weakly dependent setting (e.g. \citeauthor{Cattaneo2022}, \citeyear{Cattaneo2022}) to devise test statistics that control size uniformly over the evaluation point $x$. An example is the construction of uniform-in-$x$ confidence bands for local polynomial quantile regression estimators with time series data. We intend to study such procedures in future research.
	\appendix

	\section{Proof of \Cref{theorem_main}}
	\label{app_proof_main}
	We first prove a preliminary result for uniform random variables.
	\begin{lemma}
		\label{lemma_maximal}
		Let $\{(U_i, Z_i)\}_{i\in \mathbb{N}}$ be a sequence of random variables defined on a common probability space $(\Omega, \Sigma, \mathbb{P})$, with mixing coefficients satisfying, for some $c > 0$, $\alpha(i) \leq \exp(-2 c i)$ $\forall i \in \mathbb{N}$. Suppose that each $U_i$ is uniformly distributed on $[0,1]$, and that each $Z_i$ takes values on a measurable space $(\mathcal{Z}, \Sigma)$.  Let $\mathcal{F}$ be a nonnegative pointwise measurable class of real-valued functions with domain $\mathcal{Z}$. Define:
		
		$$Z_n(u,f) \coloneqq \frac{1}{\sqrt{n}}\left(\sum_{i=1}^n \mathbf{1}\{U_i \leq u\} f(Z_i) - \mathbb{E}[\mathbf{1}\{U_i \leq u\}f(Z_i)] \right) \, , \quad u \in [0,1],  f \in \mathcal{F} \, .$$
		
		Suppose that the class of functions $\mathcal{F}$ admits a measurable and  bounded envelope $\kappa$. Assume that, for some $q \in (1,\infty]$, $\omega(\epsilon) \coloneqq \sup_{P \in \mathcal{P}(\mathcal{Z})}\mathcal{N}(2\epsilon \lVert \kappa \rVert_\infty,\mathcal{F}, \lVert \cdot \rVert_{q,P})$, is finite for every $\epsilon > 0$, where $\lVert f \rVert_{P,q} = (P|f|^q)^{1/q}$ and the supremum is taken over all probability measures on $(\mathcal{Z}, \Sigma)$. Then, for every $n \geq 2$ and  $K \in \mathbb{N}$ such that $K \leq \lceil \log_2(n) \rceil$:
		
		$$\left\lVert \sup_{u \in [0,1], f \in \mathcal{F} }|Z_n(u,f) - Z_n(\Pi_K(u),f)| \right\rVert_{\psi_1}\leq \frac{\lVert \kappa \rVert_\infty}{\sqrt{n}} + D \lVert \kappa\rVert_\infty\sum_{l=K+1}^{\lceil \log_2(n) \rceil}  \inf_{\delta \geq 0} \psi_l(\delta) \, ,$$
		where the constant $D > 0$ depends solely on $c$; for $J \in \mathbb{N}$, $\Pi_J(u) = \frac{\lfloor 2^J u \rfloor}{2^J}$; and:
		\begin{equation*}
			\begin{aligned}
						\psi_l(\delta) = \left(n^{-1/2}\log(n)^2[(l+1)\lor \log(\omega(\delta))] + \sqrt{\frac{l}{2^l} \lor \frac{1}{n}}\sqrt{[(l+1)\lor \log(\omega(\delta))]}\right) + \\
						 \sqrt{n}\delta \left[\left(n^{-1}\log(n)^2(l+1) + \sqrt{\frac{l}{2^l n} \lor \frac{1}{n^2}}\sqrt{(l+1)}\right)^\frac{{q-1}}{q} +\frac{1}{2^{\frac{(q-1)}{q}(l+1)}}\right] \, .
			\end{aligned}
		\end{equation*}
		
		\begin{proof}			
			We begin by adapting the chaining argument in the proof of Proposition 7.2. of \cite{Rio2017}. Fix $f \in \mathcal{F}$. Observe that for $L \geq K$, we may write:
			
			$$\sup_{u \in [0,1]} | Z_n(u,f) - Z_n(\Pi_K(u), f)| \leq \sum_{l= K+1}^{L} \Delta_l(f) + \Delta^*_L(f) \, , $$
			where $\Delta_l(f) = \max_{k \in \{1, \ldots, 2^l\}} |Z_n((k-1)/2^l,f) - Z_n(k/2^l,f)| $; and $\Delta^*_L(f) = \sup_{u \in [0,1]} |Z_n(\Pi_L(u),f) - Z_n(\Pi(u), f)|$. Observe that:
			
			$$ -  \lVert \kappa \rVert_{\infty}\frac{\sqrt{n}}{2^{L}} \leq Z_n(\Pi(u)) - Z_n(\Pi_L(u)) \leq \Delta_{L}(f)  + \lVert \kappa \rVert_{\infty}\frac{\sqrt{n}}{2^{L}} \,.$$

			From the above, we obtain that:
			
			$$\sup_{u \in [0,1]} | Z_n(u,f) - Z_n(\Pi_K(u), f)| \leq 2 \sum_{l=K+1}^L \Delta_l(f) + \lVert \kappa \rVert_{\infty} \frac{\sqrt{n}}{2^L}  \, ,$$
			and, by taking $L = \lceil\log_2(n)\rceil$, we arrive at:
			
			\begin{equation}
				\label{eq_terms}
				\begin{aligned}		
					\sup_{u \in [0,1]} | Z_n(u,f) - Z_n(\Pi_K(u), f)| \leq 2 \sum_{l=K+1}^{\lceil \log_2(n) \rceil} \Delta_l(f) + \lVert \kappa \rVert_{\infty} \frac{1}{\sqrt{n}} \, .
				\end{aligned}
			\end{equation}
			
			The previous argument holds for a fixed $f$. We achieve uniformity as follows. For a given sequence $\{\delta_l\}_{l=K+1}^{\lceil \log_2(n)\rceil}$, we may write, for each $l$:
			
			\begin{equation}
				\begin{aligned}	
					\label{eq_decomposition}		
				\sup_{f \in \mathcal{F}}	\Delta_l(f) \leq \max_{f \in T_l} \max_{k \in \{1, \ldots, 2^l\}} |Z_n((k-1)/2^l,f) - Z_n(k/2^l,f)|  \\ + \sup_{f \in \mathcal{F}: \lVert f - f'\rVert_{q,\hat{\mathbb{P}}_n} \lor \lVert f - f'\rVert_{q,\bar{\mathbb{P}}_n} \leq \delta_l}  \max_{k \in \{1, \ldots, 2^l\}} |Z_n((k-1)/2^l,f) - Z_n(k/2^l,f) - (Z_n((k-1)/2^l,f') - Z_n(k/2^l,f'))|  \, ,
				\end{aligned}
			\end{equation}
			with $T_l$ being, simultaneously, a $2\delta_l \lVert \kappa \rVert_\infty$-cover of $\mathcal{F}$ with respect to $\lVert \cdot \rVert_{q,\bar{\mathbb{P}}_n}$  and $\lVert \cdot \rVert_{q,\hat{\mathbb{P}}_n}$, where $\lVert \kappa \rVert_{q,\bar{\mathbb{P}}_n}= \left(\frac{1}{n}\sum_{i=1}^n \mathbb{E}[|f(Z_i)|^q]\right)^{1/q}$; and $\hat{\mathbb{P}}_n$ denotes the empirical measure $\frac{1}{n}\sum_{i=1}^n \delta_{Z_i}$. Observe that, by the assumption in the statement of the lemma, $|T_l| \leq \omega(\delta_l)^2$; and that the set $T_l$ is random. By filling the $T_l$ with constant functions equal to zero whenever $|T_l| < \omega(\delta_l)^2$, we may assume, without loss of generality, that $|T_l| = \omega(\delta_l)^2$ and we may thus order the random elements in $T_l = \{f_{1,l}, f_{2,l}, \ldots, f_{\omega(\delta_l)^2, l}\}$. We thus consider $T_l$ to be a set of $\omega(\delta_l)^2$ random functions.
			
			We will deal with each term on the right-hand side of \eqref{eq_decomposition} separately. Consider the first term. For a given $l \in \mathbb{N}$ and some $k \in \{1,\ldots, 2^l\}$ and $f \in T_l$, define $Y_{i,l}(k,f) = \frac{1}{\sqrt{n}}f(Z_i)\mathbf{1}\{ (k-1)/2^l \leq U_i \leq k/2^l \}$. Rio's covariance inequality \citep[Theorem 1.1]{Rio2017} yields:
			
			$$\sum_{s\geq i}^\infty |\mathbb{C}(Y_{i,l}(j,f), Y_{s,l}(j,f))| \leq \frac{4 \lVert \kappa \rVert_\infty^2}{n} \sum_{s=0}^\infty  \alpha(s) \land \frac{1}{2^l} \leq  \frac{4 \lVert \kappa \rVert_\infty^2}{n}  \left(\frac{l \log(2)}{c2}  \frac{1}{2^l} + \frac{1}{2^{l-1}}\frac{1}{1 - \exp(-2c)} \right) \eqqcolon v_l^2 n^{-1} \lVert \kappa \rVert^2 \, . $$
			
			It then follows by Theorem 2 of \cite{Merlevede2009} that, fo a constant $C$ depending only on $c$, we have that, for any $n \geq 2$ and every $l \in \mathbb{N}$, $k \in \{1,2,\ldots, 2^l\}$ and $f \in T_l$:
			
			\begin{equation*}
				\begin{aligned}	
				\mathbb{P}[|Z_n((k-1)/2^l,f) - Z_n(k/2^l,f)| \geq x] \leq \exp\left(\frac{Cx^2}{v^2_l  \lVert \kappa\rVert_{\infty}^2  + \lVert \kappa \rVert^2_\infty n^{-1} + x \lVert \kappa\rVert_\infty n^{-1/2} \log(n)^2} \right), \quad \forall x \geq 0 \, .
			\end{aligned}
			\end{equation*} 
			
			By Lemma 2.2.10 of \cite{vanderVaart1996}, there is a constant $K$ depending only on $c$ such that, for every $l \in \mathbb{N}$ and $n \geq 2$:
			
			\begin{equation}
				\label{eq_bound_first}
				\begin{aligned}
					\left\lVert \max_{f \in T_l} \max_{k \in \{1, \ldots, 2^l\}} |Z_n((k-1)/2^l,f) - Z_n(k/2^l,f)| \right\rVert_{\psi_1} \leq \\ K \lVert \kappa \rVert_\infty( n^{-1/2} \log(n)^2 \log(1+ 2^l\omega(\delta_l)^2) + \sqrt{ v_l^2 + n^{-1}}\sqrt{\log(1+ 2^l\omega(\delta_l)^2)}) \, .
				\end{aligned}
			\end{equation}
			
			Next, we deal with the second term in \eqref{eq_decomposition}. A simple argument reveals that:
			\begin{equation*}
				\begin{aligned}			
			\max_{k \in \{1,\ldots, 2^l\}}	\sup_{f \in \mathcal{F}: \lVert f - f'\rVert_{q,\hat{\mathbb{P}}_n} \lor \lVert f - f'\rVert_{q,\bar{\mathbb{P}}_n} \leq \delta_l} |Z_n((k-1)/2^l,f) - Z_n(k/2^l,f) - (Z_n((k-1)/2^l,f') - Z_n(k/2^l,f'))| \leq \\ 2\lVert \kappa \rVert_\infty \delta_l n^{(2 -q)/2q} \max_{k \in \{1,\ldots 2^l\}}\left|\sum_{i=1}^n \mathbf{1}\{(k-1)/2^l \leq U_i \leq k/2^l\}  \right|^{\frac{q-1}{q}}+ 2\lVert \kappa \rVert_\infty\delta_l\frac{\sqrt{n}}{2^{\left(\frac{q-1}{q}\right)l}} \, ,
				\end{aligned}
			\end{equation*}
			from which it follows that:
			
				\begin{equation}
					\label{eq_bound_second}
				\begin{aligned}			
				\left\lVert \max_{k \in \{1,\ldots, 2^l\}}	\sup_{f \in \mathcal{F}: \lVert f - f'\rVert_{q,\hat{\mathbb{P}}_n} \lor \lVert f - f'\rVert_{q,\bar{\mathbb{P}}_n} \leq \delta_l} |Z_n((k-1)/2^l,f) - Z_n(k/2^l,f) - (Z_n((k-1)/2^l,f') - Z_n(k/2^l,f'))| \right\rVert_{\psi_1} \leq \\
	2	S \lVert \kappa \rVert_\infty \delta_l  n^{1/2q} \left\lVert \max_{k \in \{1,\ldots 2^l\}} \left|\frac{1}{\sqrt{n}}\sum_{i=1}^n (\mathbf{1}\{(k-1)/2^l \leq U_i \leq k/2^l\} - 1/2^l) \right|\right\rVert_{\psi_1}^{\frac{q-1}{q}} + S 4\lVert \kappa \rVert_\infty\delta_l\frac{\sqrt{n}}{2^{\left(\frac{q-1}{q}\right)l}} \leq \\
		2	S\delta_l \lVert \kappa \rVert_\infty \sqrt{n} \left(K( n^{-1} \log(n)^2 \log(1+ 2^l) + \sqrt{n^{-1}(v_l^2 +  n^{-1})}\sqrt{\log(1+ 2^l)})^{\frac{q-1}{q}} +  \frac{2}{2^{\left(\frac{q-1}{q}\right)l}}  \right) \, ,
				\end{aligned}
			\end{equation}
			where  $S>0$ is an absolute constant, and the last inequality follows from replicating the preceding argument replacing $T_l$ with a constant function equal to one. Using that $\log(1+2^l \omega(\delta_l)^2) \leq \log(2^{l+1} \omega(\delta_l)^2) = (l+1) \log(2) + 2 \log(\omega(\delta_l))$ and $v^2_l \leq M \frac{l}{2^l}$ for a constant $M$ depending only on $c$, we can simplify both \eqref{eq_bound_first} and \eqref{eq_bound_second}. Then, plugging these back onto \eqref{eq_decomposition}, optimising, and then plugging back onto \eqref{eq_terms}; we obtain the desired result.
		\end{proof}
	\end{lemma}
		We are now ready to prove the main result. We begin by proving the result for a nonnegative class of functions with bounded envelope. Start by noticing that:
		
		\begin{equation}
			\label{eq_sum_terms}
			\begin{aligned}
							|S(x,f;h)- \mathbb{E}[S(x,f;h)]| \leq \left|\frac{1}{\sqrt{nh}}\sum_{i=1}^n (\mathbf{1}\{ x \leq X_i \leq x + h\} f(Z_i)- \mathbb{E}[\mathbf{1}\{ x \leq X_i \leq x + h\} f(Z_i)])\right| + \\ \left|\frac{1}{\sqrt{nh}}\sum_{i=1}^n (\mathbf{1}\{ x - h \leq X_i \leq x\} f(Z_i) - \mathbb{E}[\mathbf{1}\{ x - h \leq X_i \leq x \} f(Z_i)])\right| + \\
							\left| \frac{1}{\sqrt{nh}}\sum_{i=1}^n \mathbf{1}\{X_i = x\}\right| \, .
			\end{aligned}		
		\end{equation}
		
		We deal with each term separately. Consider the first term. Observe that, since each $X_i$ is continuously distributed, we have that:
		\begin{equation}
			\label{eq_uniform_rep}
			\begin{aligned}				
				\left|\frac{1}{\sqrt{nh}}\sum_{i=1}^n (\mathbf{1}\{ x \leq X_i \leq x + h\} f(Z_i)- \mathbb{E}[\mathbf{1}\{ x \leq X_i \leq x + h\} f(Z_i)])\right| = \\ \left|\frac{1}{\sqrt{nh}}\sum_{i=1}^n (\mathbf{1}\{ G(x) < U_i \leq G(x + h)\} f(Z_i)- \mathbb{E}[\mathbf{1}\{ G(x) < U_i \leq G(x + h)\} f(Z_i)])\right| + \\
				\left|\frac{1}{\sqrt{nh}}\sum_{i=1}^n \mathbf{1}\{ U_i = G(x)\} f(Z_i)\right| \, ,
			\end{aligned}
		\end{equation}
		where each $U_i \coloneqq G(X_i)$ is uniformly distributed. The last term on the right hand side of \eqref{eq_uniform_rep} is, with probability one, bounded above by $\frac{\lVert g\rVert_\infty }{\sqrt{a_n n}}$ uniformly over $h$, $f$ and $x$.  Next, note that, for any $x \in \mathbb{R}$ and $h > 0$, trivially:
		
		$$G(x+h) - G(x) \leq \lVert g\rVert_\infty h \, .$$
		
		Combining this fact with continuity of $G$ reveals that, for each $x \in \mathbb{R}$, the set:
		
		$$\Xi_x = \{G(x+h) - G(x): 0 \leq h < b_n\} \, ,$$
		is an interval contained in 
		
		$$[0, \lVert g \rVert b_n)\, .$$
		
		Therefore, setting $K = \lfloor - \log_2(\lVert g \rVert_\infty b_n)\rfloor$ yields that:
		
		\begin{equation*}
			\begin{aligned}
				\sup_{x \in \mathbb{R}} \sup_{a_n \leq h < b_n} \sup_{f \in \mathcal{F}}\left|\frac{1}{\sqrt{nh}}\sum_{i=1}^n (\mathbf{1}\{ x \leq X_i \leq x + h\} f(Z_i)- \mathbb{E}[\mathbf{1}\{ x \leq X_i \leq x + h\} f(Z_i)])\right|  \leq \\
				\frac{1}{\sqrt{a_n}}	\sup_{u \in [0,1]}\sup_{s \in [0,{2}^{-K}) }	\sup_{f \in \mathcal{F}}	\left|\frac{1}{\sqrt{n}}\sum_{i=1}^n (\mathbf{1}\{ u < U_i \leq u+s\} f(Z_i)- \mathbb{E}[\mathbf{1}\{ u < U_i \leq u + s\} f(Z_i)])\right| + \frac{\lVert g\rVert_\infty }{\sqrt{a_n n}} \, .
			\end{aligned}
		\end{equation*}
		
		 But we then note that:
		\begin{equation}
			\label{eq_bound_uniforms}
			\footnotesize
			\begin{aligned}
			\sup_{u \in  [0,1]}\sup_{s \in [0,{2}^{-K}) }	\sup_{f \in \mathcal{F}}	\left|\frac{1}{\sqrt{n}}\sum_{i=1}^n (\mathbf{1}\{ u < U_i \leq u+s\} f(Z_i)- \mathbb{E}[\mathbf{1}\{ u < U_i \leq u + s\} f(Z_i)])\right| \leq \\
			\sup_{k \in \{1,2, \ldots 2^K\}}\sup_{s \in [0,{2}^{-K}) }\sup_{f \in \mathcal{F} }	\left|\frac{1}{\sqrt{n}}\sum_{i=1}^n \left(\mathbf{1}\left\{ \frac{(k-1)}{2^K} < U_i \leq \frac{(k-1)}{2^K}+ s\right\} f(Z_i)- \mathbb{E}\left[\mathbf{1}\left\{ \frac{(k-1)}{2^K} < U_i \leq \frac{(k-1)}{2^K}+ s\right\} f(Z_i)\right]\right)\right| + \\
			\sup_{k \in \{1,2, \ldots 2^K\}}\sup_{s \in [0,{2}^{-K}) }\sup_{f \in \mathcal{F} }	\left|\frac{1}{\sqrt{n}}\sum_{i=1}^n \left(\mathbf{1}\left\{ \frac{(k-1)}{2^K} -s < U_i \leq \frac{(k-1)}{2^K}\right\} f(Z_i)- \mathbb{E}\left[\mathbf{1}\left\{ \frac{(k-1)}{2^K}-s < U_i \leq \frac{(k-1)}{2^K}\right\} f(Z_i)\right]\right)\right| \, .			
			\end{aligned}
		\end{equation}
		and, since both $U_i$ and $1-U_i$ are uniformly distributed, Lemma \ref{lemma_maximal} provides control over the Orlicz norm of both terms on the right-hand side of \eqref{eq_bound_uniforms}. An analogous argument enables controls of the uniform error of the second term in \eqref{eq_sum_terms}. Finally, as for the third term in \eqref{eq_sum_terms}, we note that its uniform error is bounded above by $\frac{\lVert \kappa \rVert_\infty}{\sqrt{na_n}}$. Applying Lemma \ref{lemma_maximal} and combining the resulting terms proves Theorem \ref{theorem_main} for nonnegative classes. 
		
		To extend the result to general bounded classes, consider the constant function $\underline{f}(z) = -\lVert \kappa\rVert_\infty$ for all $z \in \mathcal{Z}$. Next, decompose the function class as $\mathcal{F} = (\mathcal{F} - \underline{f}) + \{\underline{f}_n\}$. The class $(\mathcal{F} - \underline{f}) $ is nonnegative, admits bounded envelope $2\lVert\kappa \rVert$ and same covering numbers as $\mathcal{F}$. Therefore, the previous result applies to it. The second class of functions is a singleton and $\underline{f}_n$ has constant sign; therefore the previous result also applies to it, with $w_n(\delta) = 1$ and hence $\delta = 0$. Combining with the bound on $(\mathcal{F} - \underline{f}_n)$ yields the desired result.

	\bibliographystyle{chicago}
	\bibliography{bib}
\end{document}